\documentclass[english,twocolumn,aps,prb,longbibliography]{revtex4-2}
\newcommand{\be}{\begin{equation}}
	\newcommand{\ee}{\end{equation}}
\newcommand{\bea}{\begin{eqnarray}}
	\newcommand{\eea}{\end{eqnarray}}

\def\e{\varepsilon}

\def\m{\mu}

\def\t{\tau}

\def\s{\sigma}

\def\G{\Gamma}

\def\bk{{\bf k}}

\def\bQ{{\bf Q}}

\def\bv{{\bf v}}

\def\nn{\nonumber}
\def\lb{\label}
\def\pref#1{(\ref{#1})}

\newcount\bozza \bozza=0
\ifnum\bozza=1
\newdimen\shift \shift=-2truecm
\def\lb#1{%
	{\label{#1}\rlap{\kern\shift{$\scriptstyle#1$}}}}
\else\def\lb#1{\label{#1}} \fi

\usepackage[T1]{fontenc}
\usepackage[applemac]{inputenc}
\usepackage[english]{babel}
\usepackage{graphics}
\usepackage{amssymb}
\usepackage{amsmath}
\usepackage{overpic}
\usepackage{comment}
\usepackage{bbold}
\usepackage{siunitx}

\usepackage{natbib}

\usepackage{xcolor}
\usepackage{soul}

\graphicspath{{./Images/}}

\begin{document}
	\title{Resistivity anisotropy from the multiorbital Boltzmann equation in nematic FeSe}
	\author{Marco Marciani}
	\affiliation{Department of Physics and ISC-CNR, ``Sapienza'' University of Rome, P.le A. Moro 5, 00185 Rome, Italy}
	\author{Lara Benfatto}
	\affiliation{Department of Physics and ISC-CNR, ``Sapienza'' University of Rome, P.le A. Moro 5, 00185 Rome, Italy}

	\begin{abstract}
We compute the resistivity anisotropy in the nematic phase of FeSe from the static solution of the multiorbital Boltzmann equation. By introducing disorder at the level of the microscopic multiorbital model we show that even elastic scattering by localized impurities may lead to non-trivial anisotropic renormalization of the electronic velocities, challenging the usual understanding of transport based only on cold- and hot-spots effects. Our model takes into account both the $xz/yz$ and the recently proposed $xy$ nematic ordering. We show that the latter one has a crucial role in order to reproduce the experimentally-measured anisotropy, providing a direct fingerprint of the different nematic scenarios on the bulk transport property of FeSe.
		
	\end{abstract}
	\date{\today}

	\maketitle
	\section{Introduction}
\begin{figure}[htb!]
	\includegraphics[width=.48\textwidth]{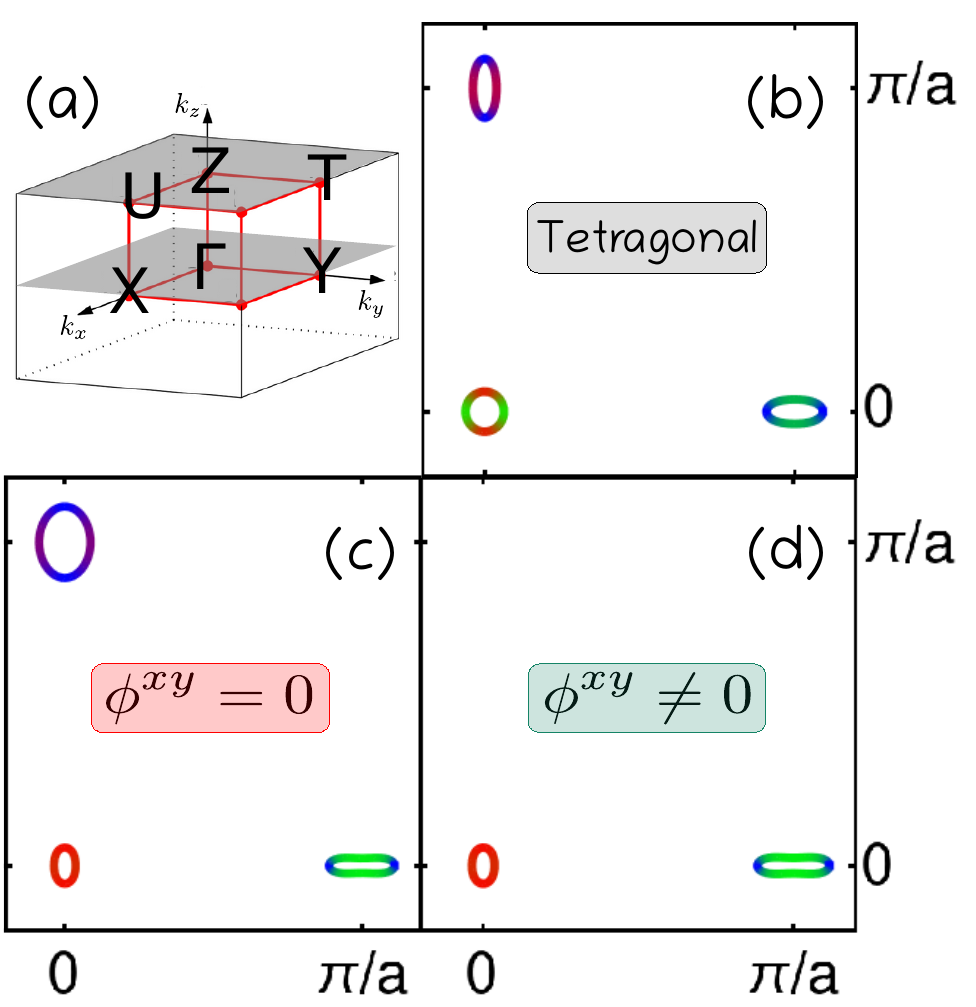}
	\caption{
	\textbf{1-Fe Brillouin Zone(BZ) in different regimes}. (a) Nomenclature of the pocket centers in the unfolded 3D BZ corresponding to 1Fe atom per unit cell; the gray planes mark the sheets at $k_z=0$ and $k_z=\pi/c$.
	(b-d) Fermi surface topology of FeSe in the 1-Fe BZ at $k_z=0$, in the normal state (b) and in the nematic phase (c,d). Panel (c) corresponds to the Fermi surface expected in the presence only of the $(xz,yz)$ nematic order parameter \pref{phixz}, while panel (d) corresponds to the Fermi surface obtained by including also the $xy$ order parameter \pref{phixy}. We assume a unique lattice spacing $a$ in the $xy$ plane.
	}
	\label{fig-FS}
	\end{figure}

Among iron-based superconductors, FeSe has a rather unique behavior, 
due to the presence of a marked nematic transition that occurs without concomitant long-range antiferromagnetic order\cite{GallaisReview16,ColdeaWatsonreview18}. In FeSe nematicity develops below the temperature $T_s=90$ K where the lattice undergoes a transition from 
tetragonal to orthorhombic structure. This metallic state is named nematic because the observed electronic anisotropy, as measured, e.g., by dc transport, is much larger than what expected from the lattice anisotropy\cite{GallaisReview16,ColdeaWatsonreview18}. In most iron pnictides the structural
transition precedes or coincides with the magnetic transition at $T_N$, below which long-range antiferromagnetic order sets in\cite{kotliar_review22}. The magnetic transition itself is generically ascribed to a nesting mechanism, favoured by the similar size among the hole-like electronic pockets at $\Gamma$ and the electron-like electronic pockets around  ${\bf Q}_X=(\pi,0)$ and  ${\bf Q}_Y=(0,\pi)$ in the 1-Fe Brillouin zone (BZ) notation, see Fig.\ \ref{fig-FS}. As a consequence, one of the earliest proposals\cite{schmalianprb12,fernandesnatphys14} identified the nematic phase as a precursor of the magnetic one, such that spins are still disordered but spin fluctuations at momentum ${\bf Q}$ break the $C_4$ lattice rotational symmetry, becoming stronger at ${\bf Q}_X$ than at ${\bf Q}_Y$.  Even though this view does not necessarily imply the existence of long-range magnetic order at a $T_N<T_s$, the lack of magnetic transition in FeSe, along with the experimental observation of a marked Fermi-surface reconstruction below $T_s$, triggered also alternative proposals, based on an orbital-ordering scenario\cite{BaekNatMat14,
SuJcondMat15, MukherjeePRL15, Jiangprb16, Chubukovprb17}. The two paradigms are actually not necessarily alternative, since also a spin-nematic scenario can lead to an effective orbital ordering once one correctly includes the orbital content of the spin fluctuations themselves, within the so-called orbital-selective spin-fluctuation scenario (OSSF)\cite{Fanfarilloprb15,Fanfarilloprb16,Fanfarillo2018}. 

From the experimental point of view the systematic investigation of the band-structure of FeSe by means of ARPES revealed a sizable deformation of the Fermi surface, which can be described via a momentum-dependent crystal-field splitting of the $d_{xz}, d_{yz}, d_{xy}$ orbitals contributing to the low-energy Fermi pockets\cite{Dingprb15,Suzuki15,Shenprb16,Fanfarilloprb16,Watsonjpsj17,Borisenkoprb18,Birgenauprx19,Kimcommphys20,watson_review22}. Above $T_s$ the Fermi surface of FeSe at $k_z=0$ consists of one circular hole-like pocket at $\Gamma$ with $xz$ and $yz$ character, and two electron-like pockets at $X$ and $Y$ with $xy$  and respectively $yz$ and $xz$ character (see Fig.\ \ref{fig-FS} panel (b)), while an additional hole-like pocket at $Z$ appears at $k_z=\pi/c$. So far, there is general consensus about the existence of a $xz/yz$ splitting that changes sign in going from the Brillouin-zone center to momenta around $\bQ_X$ or $\bQ_Y$.  This can be represented by a nematic order parameter
\be
\lb{phixz}
\Phi^{xz/yz}=\langle d^\dagger_{xz}d_{xz}-d^\dagger_{yz}d_{yz}\rangle,
\ee
that is positive at $\Gamma$ and negative at $X$ and $Y$.

On the other hand, the exact role of the $xy$ orbital is still debated. Such a debate comes along with the ongoing discussion on the presence or not of the $Y$ electron pocket below $T_s$\cite{Fanfarilloprb16,Watsonjpsj17,WatsonNJP17,Borisenkoprb18,Birgenauprx19,Kimcommphys20,watson_review22}, which is also relevant for the theoretical interpretation of the gap anisotropy observed in the superconducting state\cite{Davis2017,Hirschfeld2017,Fanfarillo2018,Rhodes2018,Chubukov2018,Si2018}. The main point is that accounting only for the $xz/yz$ splitting in Eq. \pref{phixz} a large electron pocket with mixed $yz$ and $xy$ character is expected at the $Y$ point (see Fig.\ \ref{fig-FS} panel (c)). However, such a pocket has not been resolved in the most recent ARPES measurements in detwinned samples\cite{Watsonjpsj17,Birgenauprx19,Kimcommphys20}. In order to solve this puzzle\cite{watson_review22} an alternative scenario has been recently suggested in Ref.\ [\onlinecite{Ereminnpjqm21}], where the authors proposed an additional nematic order parameter accounting for the splitting of the $xy$ occupancy in the two electron pockets, i.e.
\be
\lb{phixy}
\Phi^{xy}=\langle d^\dagger_{xy,X}d_{xy,X}-d^\dagger_{xy,Y}d_{xy,Y}\rangle.
\ee
Such an order parameter is equivalent to an anisotropic hopping between the $d_{xy}$ orbitals of the nearest-neighbours atoms in the 2-Fe unit cell, which is the physical one. The main consequence of the splitting \pref{phixy} is to readily explain the progressive disappearance of the $Y$ pocket at a temperature below $T_s$ (see Fig.\ \ref{fig-FS} panel (d)), accompanied by a Lifshitz transition. 

A second striking difference among FeSe and other families of iron-based superconductors is the different sign of the resistivity anisotropy reported below $T_s$. Indeed, while in 122 compounds\cite{fisherscience10,mazin10, degiorgi10, fisher2011,
degiorgi2012, mirriprb14} the resistivity is smaller along the longer $a$ axis (corresponding to the $\Gamma-X$ direction in the 1-Fe BZ), i.e. $\Delta \rho = \rho_x-\rho_y<0$, in FeSe the opposite behavior is observed\cite{prozorovnatcomm13,Prozorovprl16}. 
Accounting for such a difference is far from being straightforward, since the dc conductivity of a multiband metal as iron pnictides is controlled by a delicate balance among Fermi velocities, density of states and scattering rates in the various pockets. In such a situation different theoretical proposals pointed out alternatively a prominent role either of the scattering-rate anisotropy\cite{schmalianprl11,uchidaprl12,uchidaprl13,davisnatphys2013,FernandesDCACprb16,Kontaniprb17} or of the Fermi-surface deformation\cite{fisherprl14,FisherDegiorgiprl15,Degiorgiprb16}. The former approach relies mainly on the calculation of the inelastic scattering rate due to the exchange of spin fluctuations, whose anisotropy is ascribed either to the spin-nematic nature of the spin fluctuations\cite{schmalianprl11,Bre14,FernandesDCACprb16} or to a secondary effect of orbital ordering\cite{Kontaniprb17}. The predominant role of the Fermi-surface deformation was instead motivated mainly by the analysis of the nematic anisotropy at finite frequency
\cite{FisherDegiorgiprl15,Degiorgiprb16,Chinottiprb17}, which involves in principle both the scattering-rate and the plasma-frequency anisotropy. Such an analysis is however rather delicate, since from one side the two quantities are unavoidably entangled by causality relations\cite{FernandesDCACprb16}, and from the other side one should definitively take into account how interactions having a predominant  interband character, as it would be the case for spin fluctuations in iron pnictides, modify the sum-rule behavior as compared to the standard case where interactions have predominant intraband character\cite{Benfattoprb11,BenfattoBoeriprb11}. 

In general, in a system such as FeSe where orbital reconstruction is much more severe than in 122 compounds, a reasonable starting point to model transport should definitively account for the Fermi-surface nematicity. A recent calculation within the OSSF scenario pointed out that in general the Fermi-surface reconstruction and the scattering-rate anisotropy give opposite contributions to the  resistivity anisotropy\cite{Fernandez2019}. In such a situation, the overall sign of the dc anisotropy is a matter of quantitative balance that requires a thoughtful calculation where all effects are properly accounted for on the same footing.

The present paper aims at achieving this goal within a simplified but yet relevant case, i.e. solving the Boltzmann transport equation in the presence only of elastic scattering processes due to impurities, but  within a full orbital  model which describes the Fermi-surface reconstruction as measured experimentally by ARPES. As compared with previous theoretical work\cite{schmalianprl11,Bre14}, which analyzed the problem within the band language, we will show that, by correctly accounting for the orbital character of the Fermi pockets, the dc conductivity computed at the level of Boltzmnann equation has a non-trivial behavior.  Indeed, as recently discussed for a generic multiorbital case in Ref.\ [\onlinecite{marcianiprb21}], while in a single-band system the transport scattering time for isotropic impurities coincides with the quasiparticle one, in a multiorbital system this is not the case. Here the multiorbital composition of the electronic bands plays a role analogous to the momentum dependence of the scattering potential for the single-band system, with two main implications. First, even elastic scattering by isotropic impurities may induce anisotropy in the observables, an effect that has not been included in previous works focused mainly on inelastic processes\cite{schmalianprl11,Bre14}. Second, the  renormalization of the current with respect to the bare band velocity, which is equivalent to include the so-called vertex corrections within the standard Kubo approach\cite{Mah00}, is finite. 
In this paper by taking advantage of the semi-analytical solution of the multiorbital problem recently provided in Ref.\ [\onlinecite{marcianiprb21}] we will compute the dc anisotropy in FeSe testing the two nematic scenarios discussed above, where either the $xy$ nematic order parameter \pref{phixy} is included or not. We will show that in both cases the velocities renormalization  due to disorder significantly contributes to the resistivity anisotropy, and becomes crucial to account for the experimental observations. More specifically, we will show that the recent proposal\cite{Ereminnpjqm21} of a $d_{xy}$ nematicity emerging along with the well-established $xz/yz$ one seems to provide a key ingredient to explain the observed resistivity anisotropy in FeSe. Our results show a direct fingerprint on a bulk material property of the $xy$ nematicity, that should be considered along with its impact on the surface ARPES probe, recently reviewed by Rhodes et al.\cite{watson_review22}. 

The plan of the paper is as follows. In Section \ref{sec:Hamilt} we introduce the low-energy orbital Hamiltonian. In Sec.\ \ref{sec:theo} we discuss the Boltzmann equation for the multiorbital model in the presence of disorder and we summarize the main results of the recent theoretical derivation\cite{marcianiprb21} of a semi-analytical solution of the integral equation for the renormalized velocities. In Sec.\ \ref{sec:FeSe} we show numerical results for FeSe in the case where both $\Phi^{xz/yz}$ and $\Phi^{xy}$ nematic order parameters are present, and we further discuss our results in Sec.\ \ref{sec:conclusions} along with the concluding remarks. The appendixes contain details of the theory and explore different parameter and disorder regimes.

	\section{The 3-orbital model} \label{sec:Hamilt}
	To describe FeSe we use a three-orbital low-energy effective model\cite{Vaf13}, as properly tailored in Ref. [\onlinecite{Ereminnpjqm21}] to fit ARPES data\cite{WatsonNJP17}. Only three spinfull $d$-orbitals are retained, whose creation operators we collect in the vector ${\bf \Psi}_\s=\left(d^{xz}_\s,d^{yz}_\s,d^{xy}_\s\right)^T$. Thus the Hamiltonian, expanded at momenta close to the pockets centers, reads as:
	\begin{equation}
	\label{h_full}
		H_T = \sum_{\bk\sigma}{\bf \Psi}^\dagger_{\bk\sigma}
		\left(H^0_{\bk,\sigma} + H^\Phi_{T}-\m_T\right)
		{\bf \Psi}_{\bk\sigma}
	\end{equation}
	where $\m$ is the chemical potential; $H^0$ contains the temperature-independent uncorrelated Hamiltonian; $H^\Phi$ accounts for the nematic deformation of the band structure and it is assumed to be independent of the local quasimomentum and spin, but dependent on the temperature $T$ through the order parameters $\Phi_T = \Phi_{0}\sqrt{1-T/T_s}$. Within the OSFF scenario\cite{Fanfarilloprb15,Fanfarilloprb16,Fanfarillo2018} such a temperature scaling arises naturally, since $H^\Phi$ encodes the real part of the nematic self-energy corrections due to exchange of spin fluctuations among hole-like and electron-like pockets. However, at the level of the present computation these can be seen as phenomenological parameters used to reproduce the ARPES data, in the same spirit of Ref. [\onlinecite{Ereminnpjqm21}]. 
	
	In FeSe the Fermi pockets are almost cylindrical in the direction perpendicular to the FeSe planes. This allows us to approximate the Fermi pockets as a stack of two cylinders with different basis whose geometrical centers are located respectively at the points $\Gamma$ and $X,Y$ and at the points $Z$ and $U,T$ of the 1-Fe Brillouin zone (BZ), see Fig. \ref{fig-FS} panel (a). Being the dispersion weakly $k_z$-dependent, we further simplify the three-dimensional (3D) BZ as the sum of two 2D BZs at $k_z=0,\frac{\pi}{c}$ and take only the cylinders bases as the relevant 2D pockets ($c$ is the lattice spacing along $z$-axis and we can assume a unique lattice spacing $a$ along the $xy$-plane also in the nematic phase). Thus, the sum $\sum_{\bk;\sigma} $ over BZ states in Eq. \eqref{h_full} due to these simplifications is equivalent to $L_z/(2c)\sum_{k_x,k_y;k_z=0,\frac{\pi}{c};\sigma}$, with $L_z$ the thickness of the sample.
	Another simplification occurs. At each pocket only two out of three spinfull orbitals contributes to the physics at the Fermi energy. Thus, at momenta close to the points $\Gamma,X,Y$ we are allowed to remove from the spinor ${\bf \Psi}$ the orbitals $d^{xy},d^{xz},d^{yz}$ respectively, and the effective Hamiltonian is described as a $4\times4$ matrix.    
	The self-energy corrections at different points of the BZ at $k_z=0$ are explicitly given by (at $k_z=\frac{\pi}{c}$ expressions are formally the same but the parameter values are different):
	\begin{eqnarray}
			H^{\Phi\,\Gamma}_{T}&=& \Phi^h_T\,\tau_3\otimes\sigma_0, \nn\\
			H^{\Phi\,X/Y}_{T}&=&  \left(\frac{\Delta\epsilon^{xy}_T}{2} \pm\frac{\Phi^{e}_{T}-\Phi^{xy}_{T}}{2}\right)\t_0\otimes\s_0 \nn\\
			&&- \left(\frac{\Delta\epsilon^{xy}_T}{2} \mp\frac{\Phi^{e}_{T}+\Phi^{xy}_{T}}{2}\right)\t_3\otimes\s_0,
		\end{eqnarray}
	where the Pauli matrices $\t$ acts on the relevant orbital space and $\s$ on the spin one. The $\Phi^{h(e)}$ parameter here corresponds to the values of $\Phi^{xz/yz}$ near $\Gamma(X$ or $Y)$, see Eq. \eqref{phixz}. 
 The most relevant new parameters introduced in Ref.\ [\onlinecite{Ereminnpjqm21}] are the nematic order parameter \pref{phixy}, $\Phi^{xy}$, and a phenomenological energy shift $\Delta \epsilon^{xy}$ of the $d^{xy}$ orbital which also sets in at the nematic transition. The latter parameter can be regarded, from a microscopically point of view, as an Hartree shift of the $d_{xy}$ orbital possibly arising from the same interactions responsible for the $\Phi^{xy}$ nematic order parameter. Both $\Phi^{xy}$ and $\Delta \epsilon^{xy}$  are responsible in general for the lowering (raising) of the band near the $X$ ($Y$) pocket, since their effects sum up at the $Y$ pocket and partly compensate at the $X$ pocket.
The full effective Hamiltonian, the symmetries analysis\cite{Vaf13} and all parameter values are presented in App. \ref{App:HamPar}.

 \begin{figure*}[t!]
 	\includegraphics[width=0.8\textwidth]{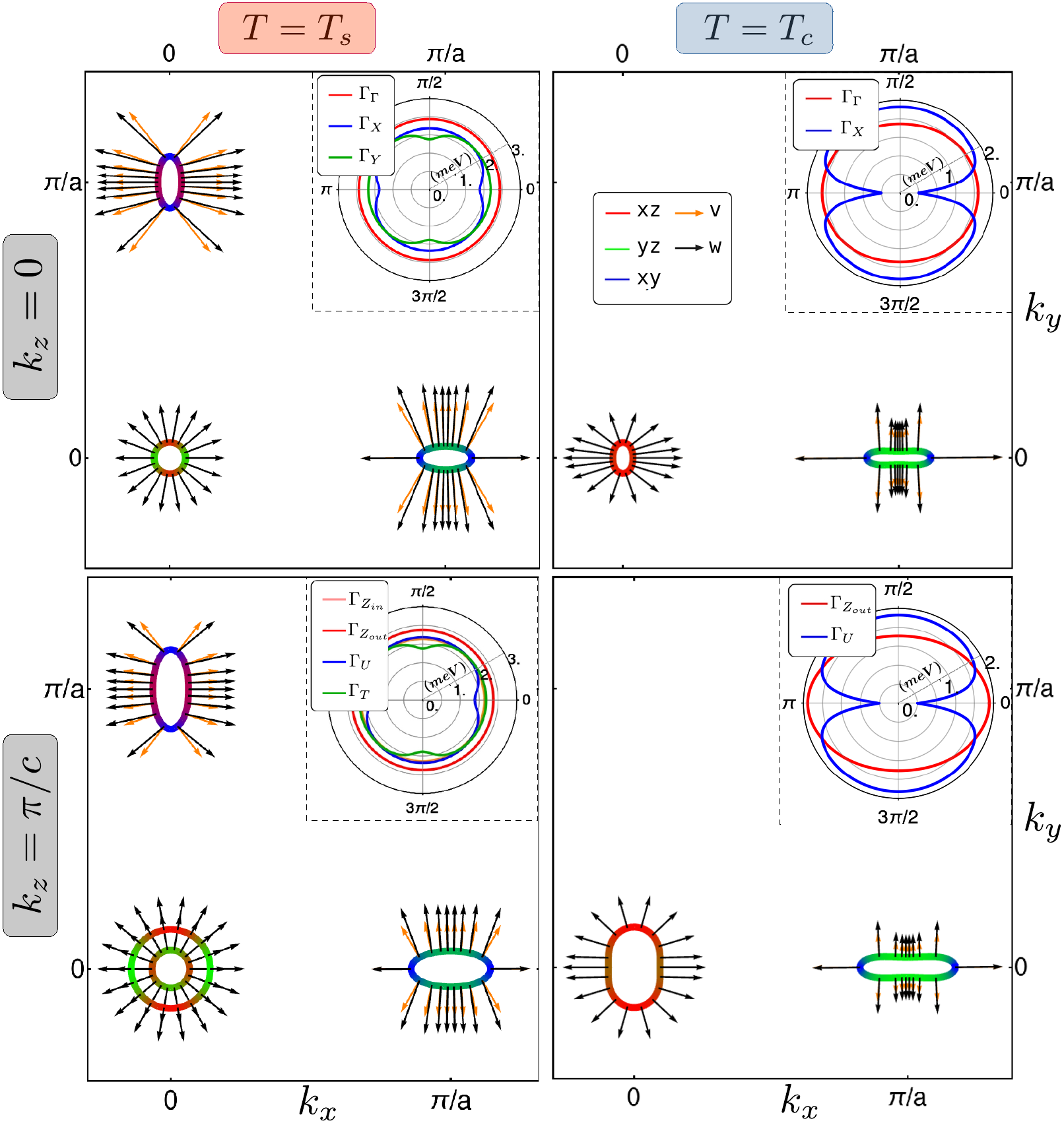}
 	\caption{
 		\textbf{Fermi pockets, velocities and rates in the 1-Fe BZ.}  BZ cut at $k_z = 0$ (top row) and at $k_z = \pi/c$ (bottom row); at the nematic phase transition $T=T_s$ (left column) and at the superconducting phase transition $T=T_c$ (right column), that represents a lower bound for our calculations. The pockets are RGB coloured according to their orbital weight. The arrows refer to the bare Fermi velocities $\bf v$ (orange line) and to the dressed ones $\bf w$ (black lines), multiplied by an overall common constant for all pockets for visualization purposes. In the insets, we show the quasiparticle rates as a function of the polar-angle coordinate along the Fermi pockets.} \label{fig:BZfull}
 \end{figure*}
 
		Lowering the temperature from $T_s$, three Lifshitz transitions take place. Right below $T_s$ there are one pocket at $\Gamma$, two pockets at $Z$ (we distinguish the inner from the outer one naming them $Z_{in}$ and $Z_{out}$), two pockets at $X$ and $Y$  and two pockets at $U$ and $T$. Lowering further the temperature the pockets at $Y$ and $T$ disappear simultaneously, at $T\simeq 70$, followed by $Z_{in}$, at $T\simeq 45$, and only four pockets are present when superconductivity sets in at $T_c=7$ K. In Fig. \ref{fig:BZfull} below we report the pocket details right at $T_s$ and at $T_c$, which represents the lower temperature bound for our calculations, which do not include superconducting effects. In the absence of the $xy$ nematic order parameter \pref{phixy} the pockets at $T_c$ are shown in Fig. \ref{fig:BZ_nophi4dexy}, for the set of parameters detailed in App. \ref{App: relevance}. In this case the $Y$ pocket survives below $T_s$ and increases in size, as a consequence of the $xz/yz$ nematicity. As we will see below, the resulting resistivity anisotropy has a completely different behavior as compared to the case when also the $xy$ nematic order parameter is present.
	
	\section{Boltzmann equation} \label{sec:theo}
		We employ the static homogeneous multiorbital Boltzmann equation to describe the dc electric transport. In its general form the equation reads as\cite{Ziman60,Sond62,Taylor1963,Wu93,Rij95,Mah00}:
		 	\begin{equation} \label{boltz_eq}
		e\, {\bf E}\cdot \nabla_{\bf k} \, \rho_{\bk,b}  =  \sum_{\bf k',b'} Q_{\bf k,b}^{\bf k',b'}(\rho_{\bk,b} - \rho_{\bk',b'}),\quad 1\leq b \leq N_{b},
	\end{equation}
	where $\rho_{\bk,b}$ is the electronic density at quasi-momentum $\bk$ and band $b$; the number of bands $N_b$ includes the spin degree of freedom. We set $\hbar=1$ in formulas.  
	We will assume that the collision kernel $Q^{bb'}_{\bf k \bf k'}$ includes only elastic scattering coming from unit-cell-localized impurities, located randomly in the sample. Such impurities affect only the local chemical potential (see  App. \ref{App: theo_meth} for details). To test the robustness of our results, other disorder types are considered in App. \ref{App: disorder}. The rates $\G_{\bk,b}$ and the lifetimes $\t_{\bk,b}$ of each state are defined as 
	\begin{equation}
	 \G_{\bk,b} = 1/\t_{\bk,b} = \sum_{\bk',b'} Q^{bb'}_{\bf k \bf k'}.   
	\end{equation} 
	To compute the dc conductivity one needs to find the change $\rho^{\bf E}_{\bk,b}$ in the distribution at linear order in the field, that can be  expressed in full generality in terms of the renormalized velocities ${\bf w}_{{\bk,b}}$ as $\rho^{\bf E}_{\bk,b} = e {\bf E} \cdot {\bf w}_{\bk,b} \,\t_{\bk,b} \,\partial_{\varepsilon^b_{\bf k}} f_{\varepsilon_{\bf k,b}}$, with $f_\varepsilon$ the Fermi function. The renormalized velocities differ in general from the bare band velocity defined as ${\bf v}^b_{\bf k}= \nabla_{\bf k}\varepsilon^b_{\bf k}$. 
    The former can be computed easily if the so-called relaxation-time approximation\cite{Mah00,Bre14,Bros16,Bri19,marcianiprb21} is implemented. This is justified whenever $\bf w$ and $\bv$ are (at least approximately) parallel, e.g., due to some symmetry of the system, with a coefficient of proportionality set by the so-called transport scattering rate.
	In most cases such as the one at hand, however, the approximation cannot be made and the computation of ${\bf w}$ has to be tackled without simplifications. In Ref.\ [\onlinecite{marcianiprb21}] we recently found an explicit solution (see App. \ref{App: theo_meth}) which is semi-analytical in the sense that it requires much less numerical computation than what would be required from a na\"ive solution strategy. As a result, the renormalized velocities can be presented as the sum of two contributions: 
	\begin{eqnarray} \label{multi_w}
		{\bf w}_{\bk,b}  &=&
		\bv_{\bk,b} + \kappa \; \sum_{mnm'n'}^{N_b}e_{\bk,b}^{m*} e_{\bk,b}^{n}  \left[(\mathbb{1} - K_{\varepsilon_\bk}) ^{-1}\right]^{mm'}_{nn'}  {\bf F}_{\varepsilon_\bk}^{m'n'}\nn\\
	\end{eqnarray}
 where $\kappa$ sets the intensity of the impurity on the scattering, $K$ is a $N_b\times N_b\times N_b\times N_b $ tensor describing the scattering among the eigenstates at the orbital level, $\bf F$ is a vector of $N_b\times N_b$ matrices where velocity and orbital content of the eigenstates are mixed (see App. \ref{App: theo_meth} for their definitions).
The second term of the equation represents the equivalent of what are usually named vertex corrections within the diagrammatic Kubo approach\cite{Mah00}.  
Finally, the conductivity is obtained as the linear response of the current density to the external field and inherits the two-contributions structure of the renormalized velocities. The first contribution is the "bare" one while the second is the correction due to the impurity scattering. They are explicitly given by:
	\begin{eqnarray} \label{sigma_cond}
		& & \quad\quad \sigma_{ij} =  \sigma^{bare}_{ij} + \sigma^{corr}_{ij} \\ \nn \\
		\sigma^{bare}_{ij} &=&  \frac{e^2}{\mathcal V}\int_\varepsilon \left(-\partial_{\varepsilon} f_{\varepsilon}\right) \sum_{b,\bk(\varepsilon)}\left( v^i\,\frac{\tau}{|v|}\,v^j \right)_{\bk,b} \nn \\
		\sigma^{corr}_{ij}  &=& \frac{e^2\,\kappa }{\mathcal V} \int_\varepsilon \; \left(-\partial_{\varepsilon} f_{\varepsilon}\right) \;\times \nn \\
		&&\sum_{mn,m'n'} \;\left(F^{i*}_{\varepsilon}  \right)^{mn} \left(\mathbb{1} - K_{\varepsilon} \right)^{-1}_{mn,m'n'}   \left(F^j_{\varepsilon}\right)^{m'n'} \nn
	\end{eqnarray}

with $e$ the electronic charge and $\mathcal{V}$ the sample volume.	
With these formulas at hand we can account for the effects of the temperature-dependent parameters $\Phi^h_T, \Phi^e_T$, $\Phi^{xy}_T$ and $\Delta \epsilon^{xy}_T$ on all the relevant quantities, i.e. the band structure, which enters via the bare velocities $\bv_{\bk,b}$, the impurity scattering, which affects both the quasiparticle scattering rates $\Gamma_{\bk,b}$ and the renormalized velocities ${\bf w}_{\bk,b}$ of each pocket, and the conductivity.

	\section{dc-conductivity anisotropy in nematic FeSe} \label{sec:FeSe}

	\subsection{Scattering rates and renormalized velocities}\label{sec:velocities}
	
	To better understand the different contributions to the dc-conductivity of nematic FeSe we show in Fig. \ref{fig:BZfull} the velocities $\bf v$, the dressed velocities $\bf w$ and the quasiparticle rates $\Gamma$ for each pocket at energy $\varepsilon=\mu_T$ for $T=T_s,T_c$. To ease the reading, in the following discussion we will refer only to the pockets at $k_z=0$, but the reader has to keep in mind that exactly the same physics takes place for the corresponding pockets at $k_z=\pi/c$. At the nematic transition the pockets at $X$ and $Y$ coincide up to a $C_4$ rotation and their dressed velocities are quite different from the bare ones. It is noticeable from the figure, however, that there are no velocity corrections along one direction, namely ${\bf w}_{x(y)}\equiv{\bf v}_{x(y))}$ for the pocket $X(Y)$. At the $\Gamma$ pocket, corrections are absent at all momenta, i.e. $\bf w \equiv \bf v$ for all states. These findings can be understood by explicit analysis of the Hamiltonian symmetries, as we detail in App. A. As far as the scattering rates are concerned, they are rather isotropic even in the electronic pockets, despite the pronounced ellipticity. We stress that the rates $\Gamma$ are small ($\sim 1\,meV$) in comparison with the bands energies ($\sim 100\,meV$), which confirms the validity of the Born approximation (to know how we determined the parameter $\kappa$ see Sec. \ref{sec:dccond}).
	
	By lowering the temperature, the $Y$ pocket sinks below the Fermi energy and the other pockets get sensibly warped. Despite these deformations, we find that a simplified approach using relaxation time approximation\cite{Mah00,marcianiprb21} would work even better at low temperatures,  since $\bf w$ and $\bf v$ are almost parallel in all pockets.  As is clear from the figure, the dressing of velocities of the $X(Y)$ pocket tends to enhance $y(x)$ conductivity. At the same time, once the $Y$ pocket has disappeared the scattering rate on the remaining $X$ pocket becomes strongly anisotropic, with cold-spots appearing in the $x$ direction. This effect can be ascribed to the lack of $xy$ orbitals in the other pockets at the Fermi surface, that results in a suppression of scattering events in the $X$ pocket at momenta where the $xy$ orbital component is the largest. As we will see below, the presence of cold-spots leads to an increase of bare conductivity along $x$, in disagreement with the experiments. However, the effect of the velocities renormalization is quantitatively larger than that of the scattering-rate suppression, and overall the $X$ pocket has enhanced conduction along $y$. Such physics may be different when a different kind of disorder is considered. For instance, with the GUE disorder considered in App. \ref{App: disorder}, the scattering rates turn out to be homogeneous across the BZ (there are no cold- or hot-spots) and there is no velocities renormalization. Still, the observed resistivity anisotropy has the right sign because in this case the bare-band velocities alone (which stay the same as those in Fig. \ref{fig:BZfull}) suffice to give a higher $y$ conductivity in the nematic phase.
 	
		\subsection{dc conductivity and resistivity anisotropy}
		\label{sec:dccond}

	\begin{figure}[t!]
		\includegraphics[width=0.45\textwidth]{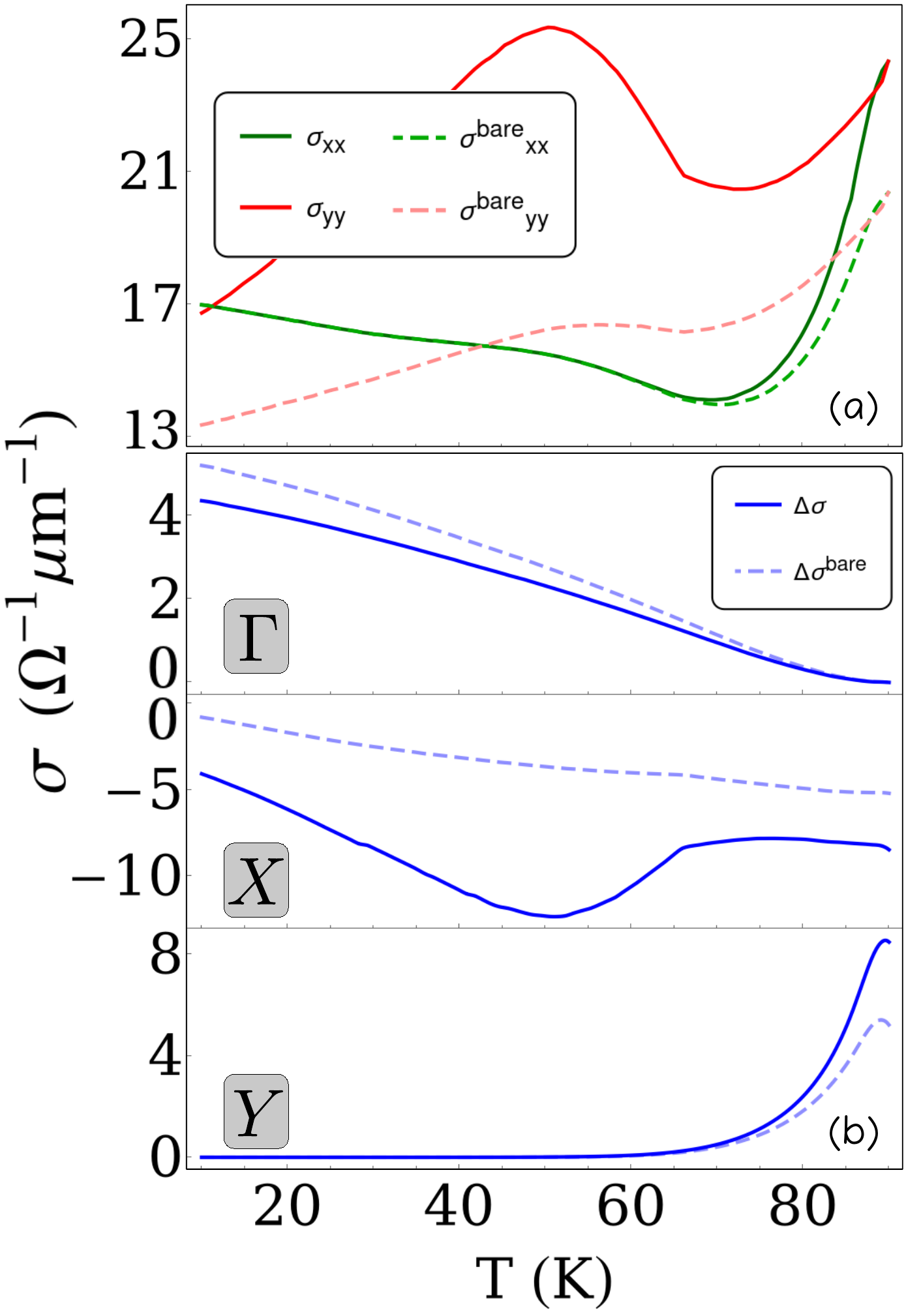}
		\caption{\textbf{Diagonal dc conductivities as a function of temperature}. (a) dc conductivity along $x$ and $y$, with (solid lines) and without (dashed lines) corrections, computed from Eq. \eqref{sigma_cond}. (b) contributions to the dc conductivity anisotropy $\Delta\sigma=\sigma_{xx}-\sigma_{yy}$ from all pockets grouped by their locations $\Gamma,X$ and $Y$ in the BZ (the contributions from the $Z,U,T$ pockets are summed to those ones respectively).}\label{fig:cond}
	\end{figure}
	
	In Fig. \ref{fig:cond} panel (a) we show the dc conductivities along $x$ and $y$. As expected from the previous considerations and the theory\cite{marcianiprb21}, $\s_{xx}<\s_{yy}$ and vertex corrections provide only positive contributions as one can see by comparing $\s$ with $\s^{bare}$. Moreover, by lowering the temperature the corrections vanish at about $65K$ for the $x$ direction as $F^{x}\simeq 0$ due to the sinking of the $Y$ pocket and a major kink appears. The different contributions of the various pockets grouped with respect to their location in the BZ are shown in Fig. \ref{fig:cond} panel (b). We must mention that the contributions from pockets lying above or below the Fermi energy and within the temperature broadening are conspicuous but qualitatively irrelevant. So, to ease the discussion, here we do not comment over their scattering rates, their velocities profiles and their contribution to the conductivities and focus only on the pockets at the Fermi level. At high temperatures the largest contributions to $\Delta \sigma\equiv \sigma_{xx}-\sigma_{yy}$ come from the $X$ and $Y$ pockets, the $\Gamma$ one being almost irrelevant (due to high scattering rates and small velocities). Decreasing the temperature the $Y$ contribution vanishes, leaving the ground to the negative $X$ contribution. Finally, at small temperatures the $\Gamma$ contribution to $\Delta \sigma$ increases, eventually beating the $X$ one. The vertex corrections of the $X$ pocket are at least two orders of magnitude bigger than the bare conductivity and make them crucial for the match with experiment. Indeed, while $\Delta \sigma^{bare}_X$ rapidly approaches zero, due mainly to the cold-spot effect mentioned earlier, the full $\Delta \sigma_X$ remains negative and compensates the positive $\Delta \sigma_\Gamma$ from the hole pocket. 
	
\begin{figure}[t!]
		\includegraphics[width=0.45\textwidth]{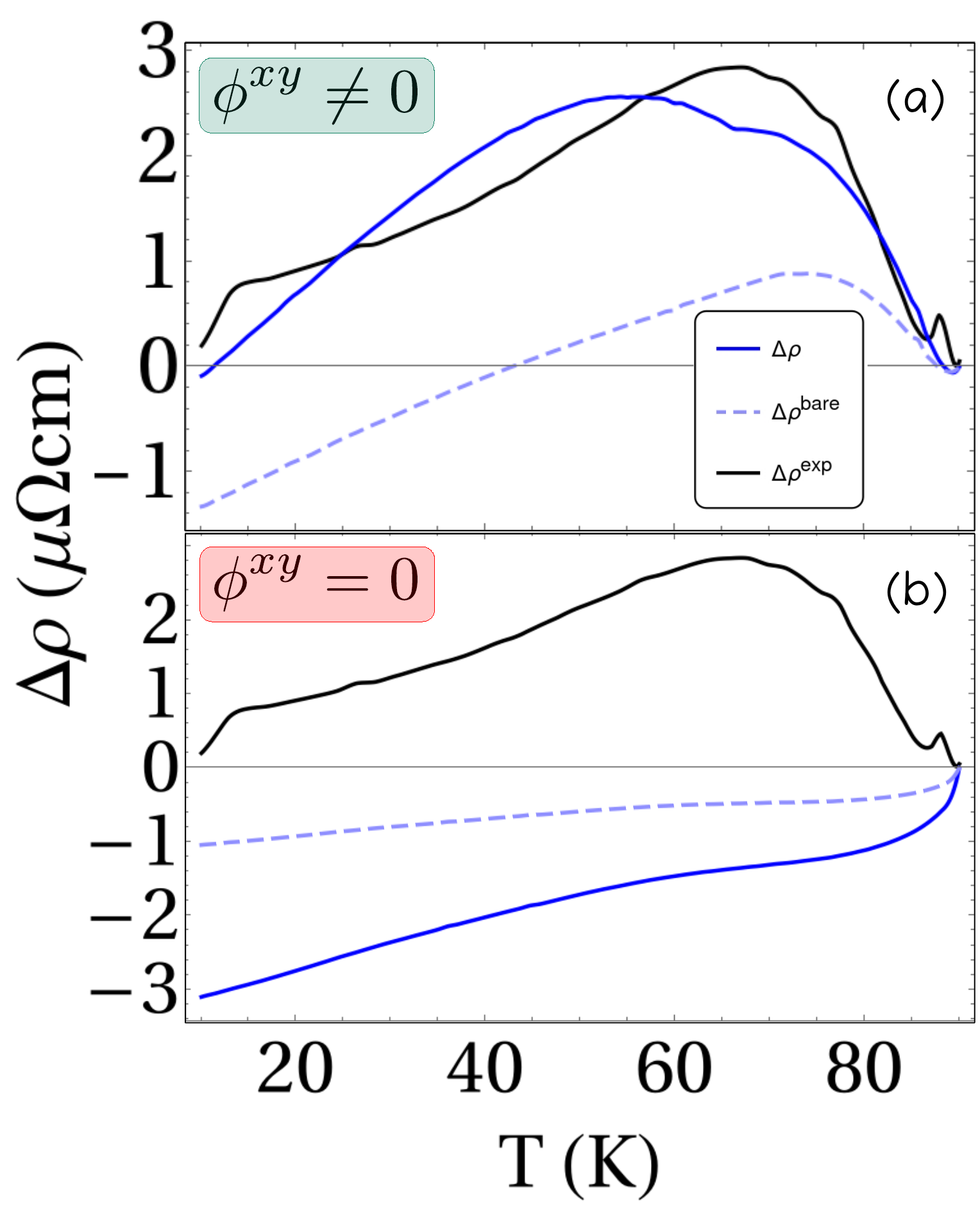}
		\caption{\textbf{Resistivity anisotropy as a function of temperature.} (a) Resistivity anisotropy obtained by including ($\Delta \rho$, blue solid line) or not ($\Delta \rho^{bare}$, blue dashed line) the velocities renormalization. We also show for comparison the experimental data ($\Delta \rho^{exp}$, black line) of the dc resistivity anisotropy taken from Ref. [\onlinecite{Prozorovprl16}]. (b) Same as panel (a) but without inclusion of the $xy$ nematic parameters  $\Phi^{xy}$ and $\Delta \epsilon^{xy}$.}   \label{fig:res}
	\end{figure}

	In Fig. \ref{fig:res} panel (a) we show the resistivity anisotropy $\Delta \rho = 1/\s_{xx} - 1/\s_{yy}$ and compare it with the experimental data of Ref. [\onlinecite{Prozorovprl16}]. The only fitting parameter is the overall scale of the resistivity controlled by the factor $n_I v_I^2$ inside $\kappa$ (see Eq.\eqref{final_multi_Q}). Since the volume of the sample of Ref. [\onlinecite{Prozorovprl16}] can be estimated to be roughly $1mm^2 \times 80\mu m$, assuming $v_I=50\,meV$ we find the impurity concentration equals $10\%$. The value is quite high considered that the experimental sample is supposed to be clean, according to the authors. However we note that on the one hand the value has a big margin of error and on the other it is already small enough to allow for the Born approximation to be reasonably good. The match between theory and experiments is very good, despite a change of sign of the theoretical curve $\Delta \rho$ at low temperature, that must be however taken with care considering that we are not including precursor effects expected before than the superconducting transition at $T_c$. The figure shows also the mentioned importance of the vertex-correction term. Clearly  $\Delta \rho^{bare}$ does not match the experiments, to the extent that it even changes sign at quite high temperatures $T\sim40K$; notice that a different $\kappa$ fitting parameter may improve the match only at high temperature, making the situation worse at lower ones.	
	We may conclude that vertex corrections (finite along $y$ and quenched along $x$) coming from the multi-orbital nature of the system are crucial in opposing the bare-bands conductivity tendency to favor negative values of $\Delta \rho$.
	
    It is also interesting to check how different nematic scenarios can affect the final results for the resistivity anisotropy. In particular, it is worth computing $\Delta \rho$ when only the $xz/yz$ nematic order parameter \pref{phixz} is introduced. Numerical details and band parameters for this case are 
discussed in App. \ref{App: relevance}.  In Fig. \ref{fig:res} panel (b) we show the final results, and one clearly sees that the resistivity anisotropy  has the wrong sign and cannot describe the experimental observations. The main reason is that when $\Phi^{xy}=0$ the $Y$ pocket survives  below $T_s$ and it even increases in size by lowering the temperature. This has the two-fold effect of suppressing the cold-spots at the $X$ pocket and, more importantly, to leave active the large conduction along $x$ due to the $Y$ pocket, where both the bare and renormalized velocities are quantitatively larger than in the $X$ pocket due to the larger size. As a consequence, vertex corrections in this case reinforce the negative trend of the bare resistivity anisotropy,  in stark contrast with the experiments. In App. \ref{App: disorder} we show a check that our theoretical model fits the experimental data also when different kinds of impurity disorder are considered.

	\section{Discussion and conclusions}{\label{sec:conclusions}}
	
	\begin{figure}[htb!]
			\includegraphics[width=.45\textwidth]{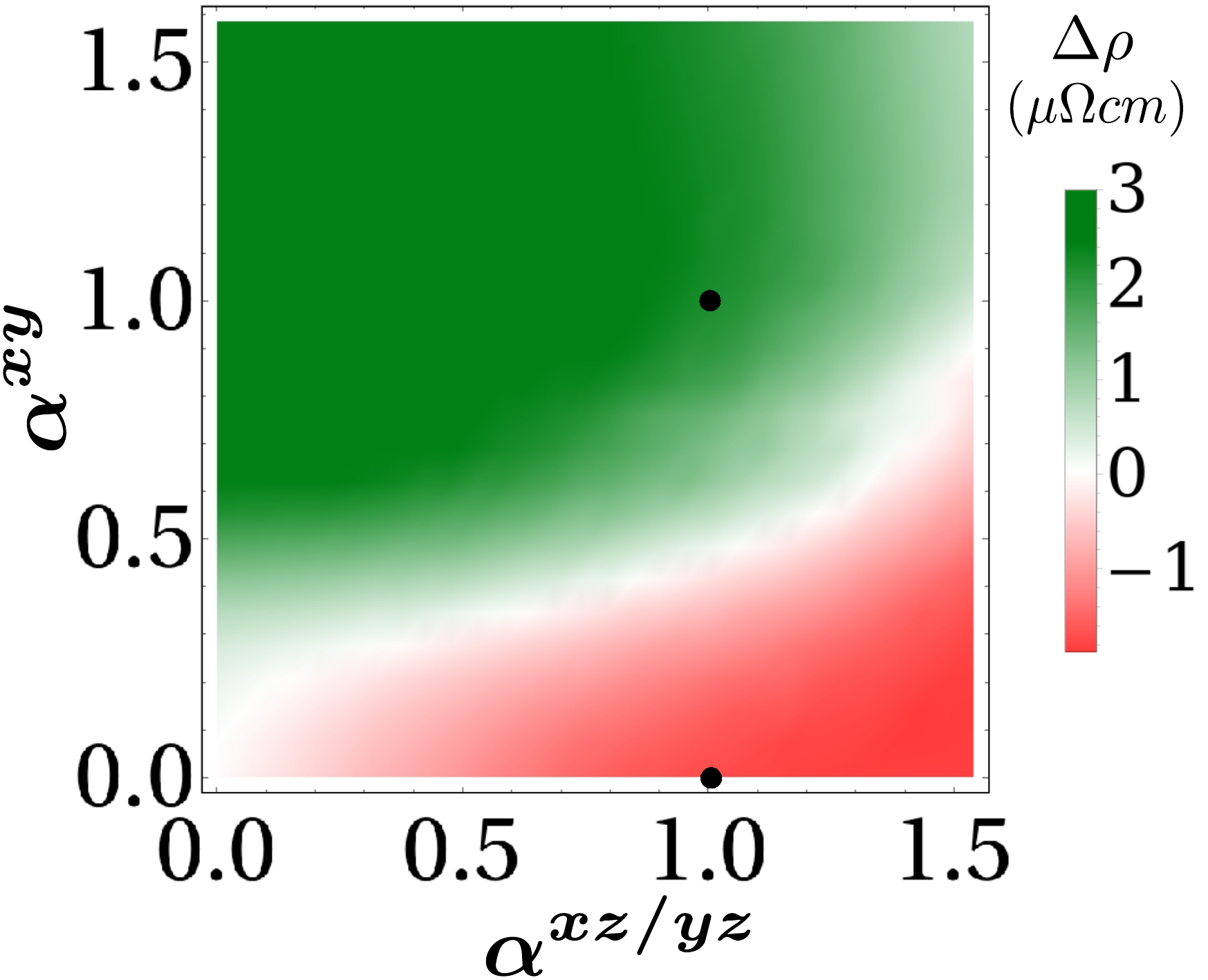}
			\caption{\textbf{Resistivity anisotropy at $T=50K$ for different values of the nematic order parameters}. For the definition of the scaling constants $\alpha_{xz/yz},\alpha_{xy}$ see the text. The chemical potential at each value of the parameters is determined self-consistently such that the electron density is the same across all values. The top and bottom black dots mark the parameter values considered in Fig. \ref{fig:res} panel (a) and (b), respectively} \label{fig:phi12vsphi4}
		\end{figure}

When comparing the results in panels (a) and (b) of Fig. \ref{fig:res} one sees that within the present approach the physical mechanism behind  the anisotropy of the transport in FeSe originates from a strong reduction of both the $xy$ and $xz$ orbital components at the Fermi surface, due to the fact that  the $Y$ pocket sinks down the Fermi level. Indeed, even though the disappearing of the $Y$ pocket induces a "cold-spot" effect, with a strong increase of the relaxation time at the electron $X$ pocket for transport along $x$, such an effect is completely overcompensated by the velocities renormalization. As a consequence, in analogy with previous findings in the context of inelastic scattering\cite{Bre14}, the usual interpretation in terms of hot and cold-spot should be taken with care, since vertex corrections actually change the final result,  and their inclusion becomes crucial to account for the experimental observations. In contrast, when the $\Phi^{xy}$ nematic order parameter is absent, as in Fig. \ref{fig:res} panel (b), vertex corrections reinforce the tendency of the bare conductivity anisotropy to favour transport along $y$, leading to an overall negative $\Delta \rho$ below $T_s$. Notice that in this view {\it both} 
the inclusion of the $xy$ nematic order parameter and the inclusion of the velocities renormalization into the dc conductivity are crucial ingredients to reproduce the resistivity anisotropy observed in FeSe.

On a wider perspective, our results suggest  that for FeSe the emergence of a $xy$ order parameter \pref{phixy} not only explains the disappearance of the $Y$ pocket in ARPES measurements\cite{watson_review22}, a surface probe, but it is also crucial to account for the resistivity anisotropy,  a bulk probe. It is worth stressing that our calculations cannot reproduce the typical linear increase of the resistivity as a function of temperature\cite{Prozorovprl16}, which is due to inelastic scattering processes that are dominant in the whole temperature range considered. However, at very low temperature (and in the absence of the superconducting transition at $T_c$) inelastic scattering would be suppressed, and the elastic one due to impurities would become dominant. In this respect we notice that the corresponding extrapolated value $\s_0^{imp}$ of the experimental conductivity approximately coincides with the average of the conductivities along $x$ and $y$ computed within our model. Since this is also the parameter setting the overall scale of the resistivity anisotropy, the present result reinforces the possibility that resistivity anisotropy can be ascribed mainly to elastic processes. Such a conclusion is further supported by various experiments carried out in annealed samples of certain electron-doped 122 compounds, where the anisotropy is suppressed with annealing\cite{uchidaprl12,uchidaprl13}. Despite the fact that electron correlations in FeSe seem to be the strongest among all Fe-based
superconductors\cite{kotliar_review22}, the fact that the parameters used reproduce both the measured Fermi surface and the extrapolated low-temperature resistivity suggests that the anisotropy could be probably linked to elastic impurity scattering also in this case.

To make a closer connection to the case of most 122 compounds, we report in  Fig.\ \ref{fig:phi12vsphi4} the resistivity anisotropy for a fixed disorder level as a function of the strength of the nematic order parameters \pref{phixz} and \pref{phixy}. To show this effect in a compact way,  we group them in two independent sets $(\Phi^{h},\Phi^{e})_{T} = \alpha_{xz/yz}\sqrt{1-\frac{T-T_c}{T_s}}\,(\Phi^{h},\Phi^{e})_{T_c}$ and $(\Phi^{xy},\Delta \epsilon^{xy})_T = \alpha_{xy}\sqrt{1-\frac{T-T_c}{T_s}}\,(\Phi^{xy},\Delta \epsilon^{xy})_{T_c}$, governed by the parameters $\alpha_{xz/yz},\alpha_{xy}$. At $\alpha_{xz/yz}=\alpha_{xy}=1$ the resistivity anisotropy is the one shown before in Fig. \ref{fig:res}. In Fig. \ref{fig:phi12vsphi4} we plot $\Delta \rho$ as an intensity plot at generic values of $\alpha_{xz/yz},\alpha_{xy}$ but fixed $T=50K$. All the red region corresponds to the case $\Delta \rho<0$, as measured in 122 systems\cite{fisherscience10,mazin10, degiorgi10, fisher2011,
degiorgi2012, mirriprb14}, and it still persists for a moderate value of the $xy$ order parameter, showing that in principle one cannot exclude a moderate $xy$ nematicity also in 122 compounds. We remind the reader, however, that the comparison with the 122 case is still done within a four-pocket model, that does not necessarily apply to the whole 122 class. For instance in  $\mathrm{BaFe_2Se_2}$ a large hole pocket at $\Gamma$ of $d_{xy}$ character is present. Even though this hole pocket is regarded to be an incoherent one and thus not very influential in the transport\cite{Nakajima2014}, in principle one should check whether the inclusion of such an incoherent $xy$ pocket change the present results.  An other interesting feature happens in K-doped 122 compounds, where the resistivity anisotropy changes sign at moderate doping\cite{prozorovnatcomm13} from negative to positive. Such change cannot be captured by our model simply by changing the intensity of the disorder, which only changes the intensity of the resistivity. As a consequence, within our picture we can understand this result as a change of nematic ordering in the various orbitals. In particular, the experimental findings are compatible with a progressive weakening of the $xz/yz$ nematicity as compared to the $xy$ one, possibly due to a weakening of spin-nematic $xz/yz$ fluctuations  within a OSSF model due to progressive suppression of quasi-nesting conditions among hole-like and electron-like pockets with doping. An increasing of $\alpha^{xy}/\alpha^{xz/yz}$ in Fig. \ref{fig:phi12vsphi4} would imply that the systems starts at one point in the red region for the undoped compound and then as hole doping increases it moves in the green one. 
We finally mention that the different sign of the anisotropy between 122 compounds and FeSe has been explained before\cite{Fernandez2019}  by means of a simplified but fully quantum approach that was including spin-fluctuations instead of impurity scattering. However, the purpose of that paper was not to compare quantitatively with the experimental data, and since the orbital model Hamiltonian was not fitting the ARPES measurements the results had a certain degree of parameter dependency. By contrast, our model is robust against parameters change (see Fig. \ref{fig:phi12vsphi4}), against the disorder model (see App. \ref{App: disorder}) and fits the experiment at a reasonable quantitative level. Clearly a combination of the two approaches may provide a more complete view on the problem.

In summary, we computed the resistivity anisotropy in the nematic phase of FeSe  due to elastic impurity scattering. We showed that when disorder is introduced at the level of the microscopic orbital model the multiorbital structure induces non-trivial effects on the transport properties, with the emergence of velocity-renormalization effects at the level of the Boltzmann transport equation. By taking advantage of our recently derived semi-analytical solution of the problem\cite{marcianiprb21}, we  computed the resistivity anisotropy in the nematic phase, comparing the results for different nematic scenarios. We find that to reproduce the experimentally-observed resistivity anisotropy of FeSe it is crucial not only to account for the full solution of Boltzmann equation, but also to add, along with the $xz/yz$ nematic order parameter, the $xy$ nematic order recently proposed in Ref.\ \cite{Ereminnpjqm21}. The latter plays indeed a crucial role in order to reverse sign of the full conductivity with respect to the bare one, due only to the Fermi-surface deformation below $T_s$. As the $xy$ nematic order parameter is suppressed the full resistivity anisotropy has instead the same sign of the bare one, and one recovers the experimental observations in 122 compounds. Our results highlight how the additional $xy$ nematic order has a strong impact on bulk transport properties of FeSe, besides the direct effect on the Fermi-surface topology, with the disappearing\cite{Ereminnpjqm21,watson_review22} of the $Y$ pocket below $T_s$. To further test the interplay among impurity scattering and nematicity it would be interesting to explore, e.g., how transport evolves under uniaxial strain\cite{coldea_prb21}, and/or in the presence of a magnetic field. Indeed, the marked anomaly of the Hall coefficient of FeSe upon entering the nematic phase rapidly disappears as nematicity softens upon, e.g., S doping\cite{coldea_prr20}, clearly suggesting a deep connection with nematic order. The relevance of velocity renormalization for the Hall effect in compensated semimetals such as pnictides has been already emphasized in previous work accounting for interband interactions due to inelastic effects\cite{fanfarillo_prl12}. It would be then very interesting to investigate if  also  impurity scattering within a multiorbital model has a similar effect, and the role played by different nematic scenarios. Finally, the microscopic justification for the $xy$ nematicity itself remains an open challenge, with crucial implications for all families of pnictides. 

\section{acknowledgements}
We acknowledge L. Fanfarillo for useful discussions. This work has been supported by PRIN 2017 No. 2017Z8TS5B, and by Sapienza University via Grant No. RM11916B56802AFE and RM120172A8CC7CC7.	
	
	\begin{appendix}

			\section{Hamiltonian parameters and symmetries} \label{App:HamPar}

		 All terms of the full effective Hamiltonian, Eq. (3) in the main text, are defined for $k_z=0$ as:
		\begin{widetext}
		\begin{eqnarray}
			H^{0\,\Gamma }(\bk)&=&\left[\epsilon^h-\frac{k^2}{2m_h}\right]\tau_0\otimes\sigma_0
			-\left[\frac{r}{2}(k_x^2-k_y^2)\right]\tau_3\otimes\sigma_0
			+rk_xk_y\tau_1\otimes\sigma_0
			+\frac{\lambda^h_{\text{SOC}}}{2}\tau_2\otimes\sigma_3,
			\nn\\
			H^{\Phi\,\Gamma}_{T}(\bk)&=& \Phi^h_T\,\tau_3\otimes\sigma_0, \nn\\
			H^{0\,X/Y}(\bk)&=& \left[\frac{k^2}{2}\left(\frac{1}{2m_1}+\frac{1}{2m_3}\right)-\frac{1}{2}\left(\epsilon^{e1}+\epsilon^{e2}\right)
			\mp\frac{1}{4}(a_1+a_3)(k_x^2-k_y^2)\right] \t_0\otimes\s_0, \nn\\
			&&+\left[\frac{k^2}{2}\left(\frac{1}{2m_1}-\frac{1}{2m_3}\right)-\frac{1}{2}\left(\epsilon^{e1}-\epsilon^{e2}\right)\mp\frac{1}{4}(a_1-a_3)(k_x^2-k_y^2)
			\right]\t_3\otimes\s_0,\nn\\
			&& +\; v_{X/Y}(\bk)\t_2\otimes\s_0, \nn\\
			\lb{hamtot}
			H^{\Phi\,X/Y}_{T}&=&  \left(\frac{\Delta\epsilon^{xy}_T}{2} \pm\frac{\Phi^{e}_{T}-\Phi^{xy}_{T}}{2}\right)\t_0\otimes\s_0 - \left(\frac{\Delta\epsilon^{xy}_T}{2} \mp\frac{\Phi^{e}_{T}+\Phi^{xy}_{T}}{2}\right)\t_3\otimes\s_0,
		\end{eqnarray}
	\end{widetext}
	where the Pauli matrices $\t,\s$ act on the orbital and spin spaces, respectively, and the slash in the label $X/Y$ is linked with $\pm$ symbols; we denote\cite{Note1}
	\begin{align}
		\nu_{X}(\bk)&=\sqrt{2}vk_y+\frac{p_1}{\sqrt{2}}\left(k_y^3+3k_yk_x^2\right)-\frac{p_2}{\sqrt{2}}k_y\left(k_x^2-k_y^2\right)\nn\\ \nu_{Y}(\bk)&=\sqrt{2}vk_x+\frac{p_1}{\sqrt{2}}\left(k_x^3+3k_xk_y^2\right)-\frac{p_2}{\sqrt{2}}k_x\left(k_y^2-k_x^2\right).
	\end{align}
At $k_z=\pi/c$ the Hamiltonian terms have the same formal expressions but different parameter values. 
The chemical potential $\mu_T$ is computed so to ensure the same average number of electrons at all temperatures taking as a reference value the one at $10K$ given in Ref. [\onlinecite{Ereminnpjqm21}].
All values of the parameters appearing in the Hamiltonian are taken from the same reference and are listed in Table \ref{Tab.FittingZ}. We use the 1-Fe lattice constant $a=2.61\,\si{\angstrom}$ and $c=5.52\, \si{\angstrom}$ (the $b$ lattice constant differs from $a$ in the nematic phase only by a tiny fraction that we neglect)\cite{Boh13}. Notice that the $\Delta \epsilon^{xy}$ and $\Phi^{xy}$ have the same effects on the $xy$ orbital at the $Y$ point. However at the $X$ one they act distructively thus being almost ineffective on this pocket.
In principle the $X,Y,U,T$ pockets experience a tiny spin-orbit coupling $\lambda^e_{SOC}=4meV$ that we neglect to simplify the numerics. The approximation involves a change in the actual Fermi surface removing the anticrossing between spin bands. Even though this change is relevant for Hall measurements, it is not for diagonal conductances due to the small energy magnitude of the spin-orbit coupling as compared to that of other parameters.

		\begin{table}[htb]
		\centering
            \renewcommand{\arraystretch}{1.2}
			\begin{tabular}{c|c c c c c}	
				\hline
				& $\Gamma$& $Z$ &&  \\ 
				\hline 
				$\epsilon^h$ &-8&$ 12 $  & &$\text{meV}$ \\ 
				$\frac{1}{2m_h}$ & $4730$&$ 1998.4 $  & &$\text{meV}\enspace\mathring{A}^2$ \\ 
				
				$r$ &$4664$&$1970.54 $ && $\text{meV}\enspace\mathring{A}^2$ \\ 
				
				$\Phi^{h}_{T}$ &$15$&$ 15 $ & &$\text{meV}$ \\ 		
				
				$\lambda^h_{\text{SOC}}$ &$23$&$23$ & &$\text{meV}$ \\ 
				\hline 		\hline 
				&$X,Y$ & $U,T$  & \\ 
				\hline 
				$\epsilon^{e1}$ &$30.6$&  $30.6$ & &$\text{meV}$ \\ 
				
				$\epsilon^{e2}$ &$48.6$ &  $48.6$ & &$\text{meV}$ \\ 
				
				$\frac{1}{2m_1}$ &$10.2060$&  $4.54$ &&$\text{meV}\enspace\mathring{A}^2$ \\ 
				
				$\frac{1}{2m_3}$ &$1355.9$&  $602.64$ &&$\text{meV}\enspace\mathring{A}^2$ \\ 
				
				$\alpha_1$ &$991.44$&  $440.64$ &&$\text{meV}\mathring{A}^2$ \\ 
				
				$\alpha_3$ &$-2937.9$&  $-1305.7$ &&$\text{meV}\enspace\mathring{A}^2$ \\ 
				
				$v$ &$-329.4$&  $-219.6$ &&$\text{meV}\enspace\mathring{A}$ \\ 
				
				$p_{z_1}$ &$-2700.9$ & $-800.27$ &&$\text{meV}\enspace\mathring{A}^3$ \\
				
				$p_{z_2}$ &$-229.7$&   $-68.06$ && $\text{meV}\enspace\mathring{A}^3$\\
			
				$\Phi^{e}_{T}$ &$-26$&  $-26$ &&$\text{meV}$\\
				
				$\Phi^{xy}_{T}$ &$45$&  $45$ && $\text{meV}$\\	
				
				$\Delta\epsilon^{xy}_T$ &$40$  & $40$ && $\text{meV}$	 
			\end{tabular} 		 
			\caption{\textbf{Hamiltonian parameters list.} Effective Hamiltonian parameters for each sector of the 1-Fe BZ, obtained by fitting ARPES data at $T=T_c$. (Adapted from Ref. [\onlinecite{Ereminnpjqm21}])\label{Tab.FittingZ}}
		\end{table}

The Hamiltonian has the following symmetries\cite{Vaf13} (the symmetries at $Z,U,T$ have the same form of those at $\Gamma,X,Y$ shown here):
 
 i) $x/y$-axis reflection symmetry at $X/Y$ and inversion symmetry at $\Gamma$ i.e. $H^{(X)}(k_x,k_y) = H^{(X)}(-k_x,k_y)$, $H^{(Y)}(k_x,k_y) = H^{(Y)}(k_x,-k_y)$ and $H^{(\Gamma )}(k_x,k_y) = H^{(\Gamma)}(-k_x,-k_y)$ (here, the origins of the momenta are the $X,Y$ and $\Gamma$ points respectively);
 
 ii)$H^{(X)}(k_x,k_y) = \tau_3 H^{(X)}(k_x,-k_y) \tau_3$ and $H^{(Y)}(k_x,k_y) = \tau_3 H^{(Y)}(-k_x,k_y) \tau_3$ where the Pauli matrix act on the orbital degree of freedom;
 
 iii) time reversal symmetry $H_{\bk} = \s_2 \left(H_{-\bk}\right)^* \s_2$, where the Pauli matrix acts on the spin degree of freedom.
 
 iv) spin-rotation invariance at $X$ and $Y$, due to the neglect of spin-orbit coupling (see App. \ref{App:HamPar}), while only conservation of the $z$-component of the spin at $\Gamma$.
 
 Notice that these symmetries imply that the up and down spin sectors are totally decoupled and degenerate.

			The peculiar properties of $\bf w$ discussed in Sec. \ref{sec:velocities} of the main paper stem directly from the symmetry of the Hamiltonian, in the following way. 
	By symmetries (i) $F^{x(y)}$ gets contributions only from the pocket $Y$($X$). The pocket $\Gamma$ do not contribute to $\bf F$ at all. Moreover $F^{x(y)}$ is strictly an off-diagonal matrix due to symmetries (ii) and the simultaneous invariance of $\t_{Y(X)}$ and sign reversal of $v^{x(y)}_{Y(X)}$ under $(k_x,k_y)\rightarrow (k_x,-k_y)$ ($(k_x,k_y)\rightarrow (-k_x,k_y)$).
	Following a similar reasoning the $K$ tensor, viewed as a matrix indexed with two-pockets labels both per raw and per columns, is actually a block matrix. In particular the $2\times2$ block concerning the $xy$-$xz$ and $xz$-$xy$ elements and the one concerning the $xy$-$yz$ and $yz$-$xy$ ones constitute two independent blocks that are uncoupled to the other elements. This block structure carries over to the tensor $(\mathbb{1}-K)^{-1}$. Thus, the property $w^x_X=v^x_X$ follows from the absence of match of $(\mathbb{1}-K)^{-1} F^x$, having only $xy$-$xz$ and $xz$-$xy$ elements, and the projector onto the eigenstates of $X$, having no components involving the orbital $xz$. The properties of $\bf w$ for the other pockets follow from similar arguments.
	
	It is quite remarkable how the general properties of the vertex corrections can be drawn from the analysis of $K$ and $\bf F$ despite the complexity of the system.
	

\section{Semi-analytical solution of the Boltzmann equation} \label{App: theo_meth}

	The collision integral kernel used in the Boltzmann equation \eqref{boltz_eq} is elastic and comes from a disorder term in the Hamiltonian which is diagonal in the band index. Such kind of disorder is arguably the simplest one may consider and is the one that is usually used in calculations. Interestingly enough, the collision integral kernel in not diagonal in the band index: 
	\be \lb{final_multi_Q}
	Q^{bb'}_{\bf k k'} = \kappa \,\left\vert{\bf e}_{\bk,b}\cdot {\bf e}_{\bk',b'}\right\vert^2 \; \delta(\varepsilon_{{\bf k},b}-\varepsilon_{{\bf k'},b'}),
	\ee
	where $\bf e_{\bk,b}$ represents an eigenvector of the multiorbital Hamiltonian, $\kappa = 2\pi n_{imp} v_{imp}^2 a^2 c /\mathcal{V}$ ($n_{imp}$ is the impurity concentration and $v_{imp}$ their potential energy) and $\varepsilon_{{\bf k},b}$ is the electron energy. The Boltzmann equation at zeroth order in $\bf E$ is trivially solved by the Fermi function $f({\varepsilon^b_{\bf k}})$.
	At first order the equation can be written as an integral equation for ${\bf w}_{\bk,b}$:
	\begin{eqnarray} \lb{w_solver}
		\sum_{\bf k',b'}\left(\delta_{\bf k\bf k'}\delta_{bb'} - Q^{bb'}_{\bf k\bf k'}\t^{b'}_{\bf k'}\right) {\bf w}^b_{\bf k'} =  {\bf v}^b_{\bf k}.
	\end{eqnarray}
	The numerical solution of Eq.\ \pref{w_solver} is rather demanding, since it requires us to invert the large matrix $1-Q\t$ in the grouped indexes $(\bk b)$ and $(\bk'b')$ in the l.h.s. of the equation \eqref{w_solver}. In Ref.\ [\onlinecite{marcianiprb21}] we recently proposed a semi-analytical solution to this problem. More specifically, we showed that thanks to the energy conservation implicit in the collision-integral kernel \pref{final_multi_Q} the matrix $1-Q\t$ becomes block-diagonal when the $\bk$ vectors are ordered in groups belonging to the same energy shell. These blocks are finite-rank, allowing one to reduce the problem to the inversion of a {\it small} matrix whose size $N^2_b$ is set by number of orbitals. The solution is given in Eq. \ref{multi_w}. As one can see, we reduced the complex problem of inverting the integral equation \pref{w_solver} to that of inverting the tensor $\mathbb{1} -K$ which, together with the matrix $F$, can be readily computed once the original multiorbital model \pref{h_full} has been diagonalized.

 The tensors $K$ and $\bf F$ are defined as 
\begin{equation} \lb{Kmat}
	(K_\varepsilon)_{mn,m'n'} = \kappa \;\sum_{b,\bk(\varepsilon)} 
	\left(e^{m} e^{n*} e^{m'*} e^{n'} \,\t\right)_{\bk,b} 
\end{equation}
and 
 \begin{equation} \lb{Fmat}
 	{\bf F}^{mn}_\varepsilon = \sum_{b,\bk(\varepsilon)} 
 	\left(e^{m} e^{n*} \,\tau\,\bv\right)_{\bk,b}.
 \end{equation}
where in both expressions we use the shorthand notation (valid in the thermodynamic limit) $\sum_{b,\bk(\varepsilon)} =\mathcal V \int \frac{\mathrm{d}^3\bk }{(2\pi)^3}\,\delta(\varepsilon_{\bk,b}-\varepsilon)$.
The full current density is defined as usual: ${\bf J} = -\frac{e}{\mathcal V}\sum_{\substack{\bk,b}} \left(\bv\,\rho\right)_{\bk,b} = \frac{e^2}{\mathcal V} \sum_{\bk,b} \bv_{\bk,b}\,\tau_{\bk,b} \,({\bf w}_{\bk,b} \cdot {\bf E}) \left(-\partial_{\varepsilon^b_{\bk}} f_{\varepsilon^b_{\bf k}}\right)$ and is linear in $\bf E$. By inserting the result \pref{multi_w} into this expression, the conductivity matrix of Eq. \eqref{sigma_cond} is obtained from the relation $ {\bf J} = \sigma {\bf E}$.

	\section{Nematic FeSe with GOE and GUE disorder} \label{App: disorder}

 		\begin{figure}[t!]
			\includegraphics[width=.45\textwidth]{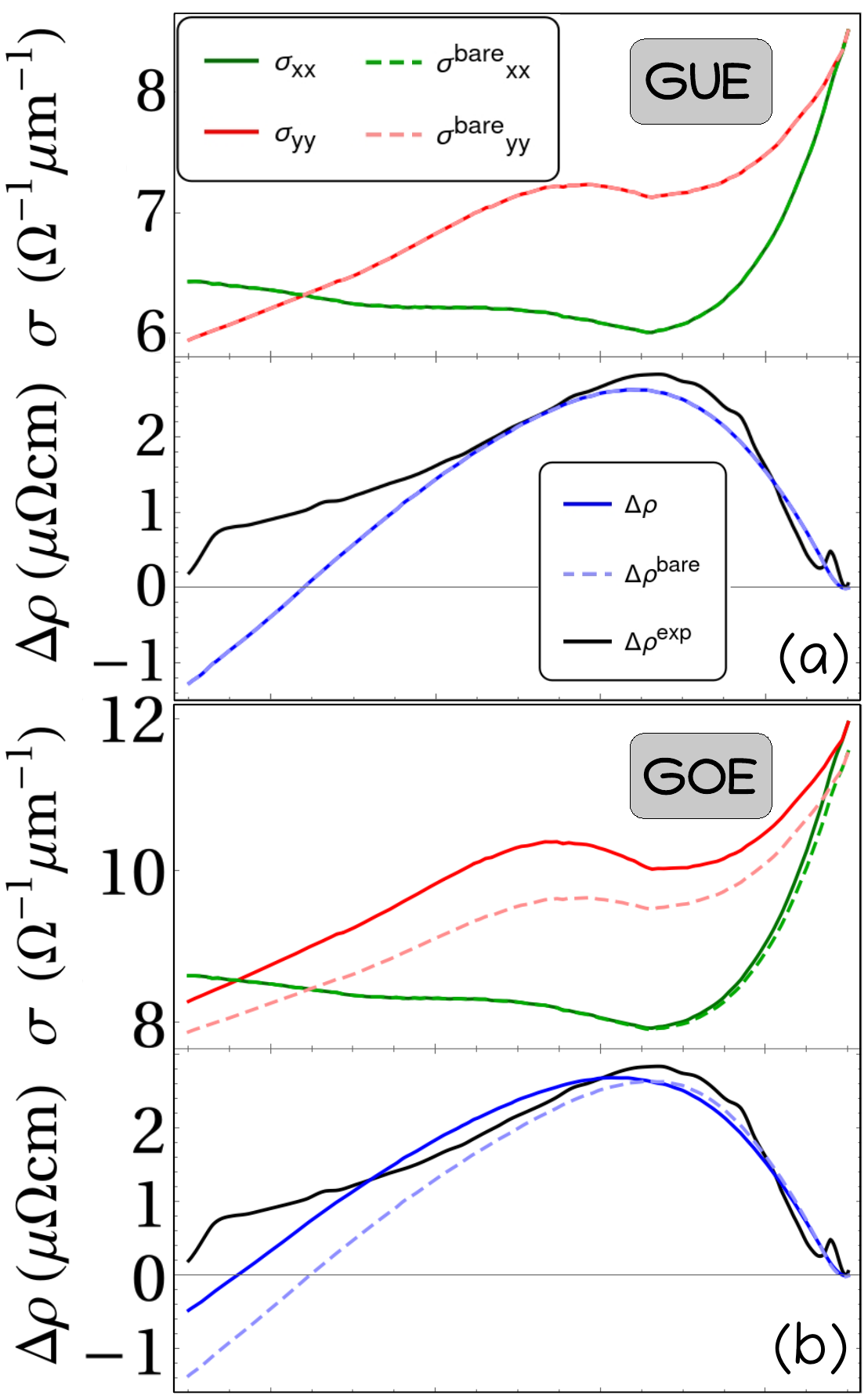}
			\caption{\textbf{Diagonal dc conductivities with different disorder types.} Conductivities along $x$ and $y$ using the same labels as in Fig. 3 and 4, assuming GUE (top) or GOE (bottom) disorder.} \label{fig:cond_res_GOE_GUE}
		\end{figure}

\begin{figure}[htb]
			\includegraphics[width=0.47\textwidth]{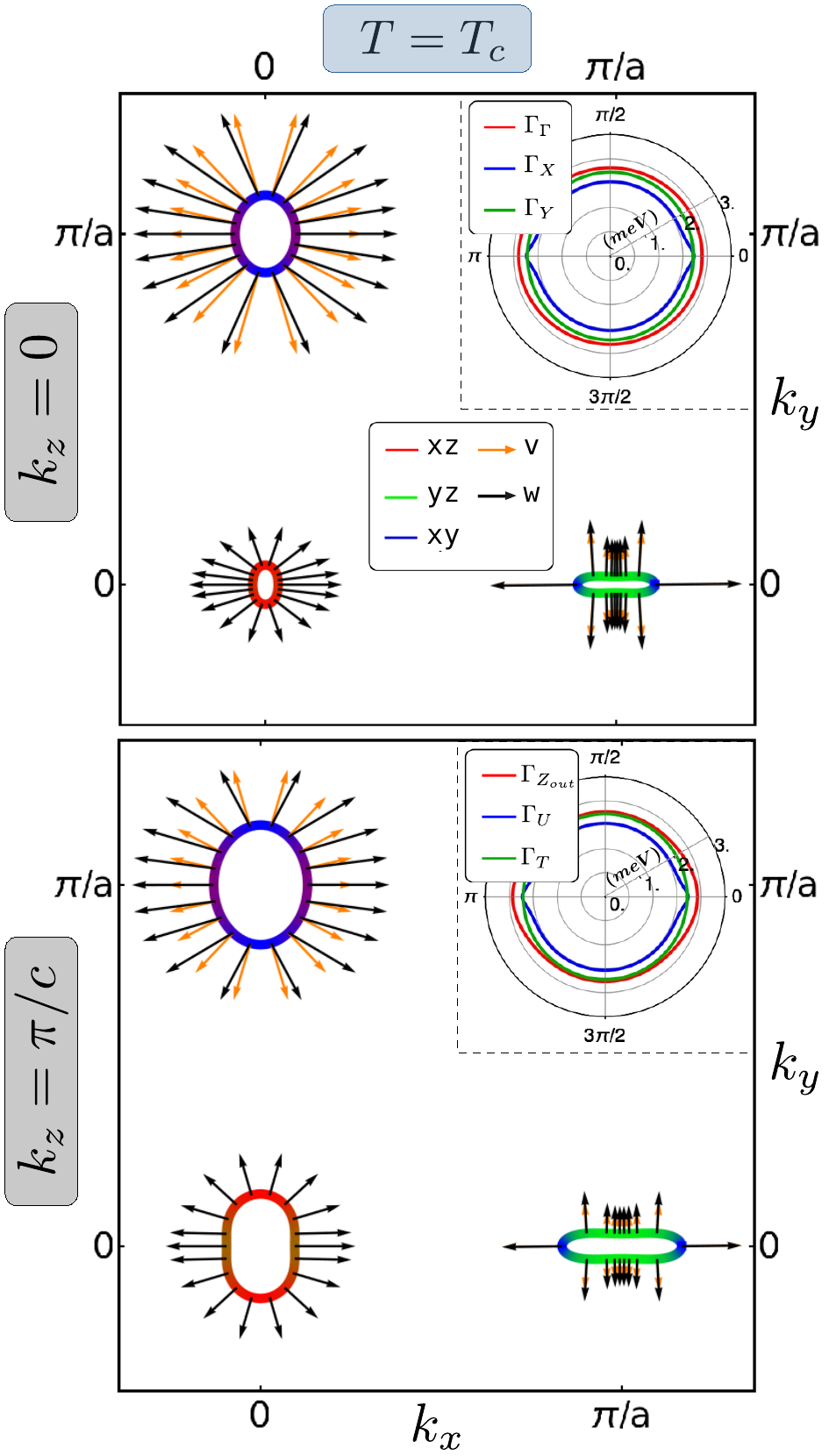}
			\caption{\textbf{Fermi pockets, velocities and rates in the 1-Fe BZ without $xy$ nematic parameters.} Same as Fig. 2, right column, but setting $\Phi^{xy}=\Delta \epsilon^{xy}=0$.} \label{fig:BZ_nophi4dexy}
		\end{figure}	
		
		Does the correct characterization of the experimentally-measured resistivity anisotropy with the model described in Sec. \ref{sec:theo} rely on the specific type of disorder chosen? In this appendix we show numerical results obtained by using different models of disorder and verify the robustness of our findings. In particular, we explore the possibility that the local impurities do not only shift the local chemical potential but may couple differently the various electronic orbitals and in a random way, so that a statistical approach would be feasible. Namely we take each impurity Hamiltonian matrix to be drawn from a Gaussian Unitary or Orthogonal Ensemble (GUE or GOE),  instead of being simple diagonal matrices of magnitude $v_{imp}$ as in the main text. The theory for such kind of disorder ensembles is slightly more complicated than the one presented in Sec. \ref{sec:theo}. The band structure and the bare velocities stay the same as before, but Eqs. (9)-(11) are different resulting in different rates and dressed velocities; we refer to Ref. [\onlinecite{marcianiprb21}] for the general formulas. In the case of FeSe, the GUE may be employed to describe disorder from impurities that produce non-vanishing microscopic magnetic fields (that vanish macroscopically) that couple the spins and break the time-reversal symmetry of $H_{eff}$. Conversely, the GOE may be employed to describe non-magnetic disorder which is diagonal in the spin space and commutes with the spinless time-reversal operator $\hat K$ (complex conjugation). The results for the conductivity and the resistivity are shown in Fig. \ref{fig:cond_res_GOE_GUE}. The fits are qualitatively similar to that of Figs.  3 panel (a) and 4 panel (a), the GUE-disordered system performing worse, the GOE one performing better. Notice how the GUE disorder induces no velocities renormalizations (dashed curves coincide exactly with the darker ones), but in this case $\Delta \rho^{bare}$ has the same sign of the experiments in most of the temperature range. This happens because there is no notion of hot- and cold-spots for this ensemble as there is a unique (isotropic) scattering rate for all states, such as in single-band models. Thus, only the bare velocities of the pockets matter which clearly favor the right sign due to the sinking of the Y pocket. The GOE disorder produces instead positive corrections, as expected from the theoretical discussion of Ref. [\onlinecite{marcianiprb21}], and also in this case such corrections are relevant in order to achieve a better agreement with experiments. These examples underline the fact that vertex corrections need not be an essential feature of the system that will describe the experiment but they must be included whenever they are finite. Finally, we remark that the absolute magnitude of the conductivities (not shown) in these cases is much smaller than in the other one, even though one tunes $\kappa$ in order to have the correct magnitude of the resistivity anisotropy. It is surprising because one could expect that the fitting parameter, that tunes separately for each disorder ensembles the matching of the resistivity anisotropy, would also create match among (the magnitudes of) the conductivities. This however is unlikely to happen, since the fitting parameter should compensate at the same time for both higher rates (roughly twice with these ensembles due to the higher number of disorder degrees of freedom) and an overall smaller relative difference between the conductivities along the $x$ and $y$ directions. We may conclude that the match between theory and experiment is quite robust against changes of the disorder type.

		\section{Dc anisotropy in the absence of $xy$ nematicity} \label{App: relevance}

To have an idea of the physics in the red region of Fig. 5, in Fig. \ref{fig:BZ_nophi4dexy} we show the Fermi surfaces at $T_c$ in the absence of the $xy$ nematic order, i.e., $\alpha^{xy}=0$ or equivalently $\Phi^{xy}_T\equiv\Delta \e^{xy}_T\equiv0$ in the Hamiltonian \pref{hamtot}. In this case the Fermi pockets at $Y$ and $T$ do not vanish, instead they increase in size as the temperature lowered. As a result, the density of states and the velocity profile favour the conductivity along the direction $x$, determining the negative resistivity anisotropy reported in Fig. 4 panel (b).

	\footnotetext[1]{We note that the sign in front of the parameter $p_2$ in the definition of $\nu_y$ differs from that in Ref. [\onlinecite{Rhodes2018}]. However with our choice $\nu_y$ respects all spatial symmetries of the system in agreement with the analysis of Ref. [\onlinecite{Vaf13}].}

	\end{appendix}
	
	\bibliography{Literature.bib}

\end{document}